\documentclass[aps, prd, amsmath, floats, floatfix, twocolumn, nofootinbib, superscriptaddress]{revtex4-1}
\usepackage{graphicx}
\usepackage{color}
\usepackage{latexsym}
\usepackage[colorlinks]{hyperref}

\newcommand{\be}{\begin{eqnarray}}
\newcommand{\ee}{\end{eqnarray}}
\newcommand{\rar}{\rightarrow}

\begin{document}

\title{Singularity-free black holes in conformal gravity: new observational constraints}

\author{Menglei~Zhou}
\affiliation{Center for Field Theory and Particle Physics and Department of Physics, Fudan University, 200438 Shanghai, China}

\author{Askar~B.~Abdikamalov}
\affiliation{Center for Field Theory and Particle Physics and Department of Physics, Fudan University, 200438 Shanghai, China}

\author{Dimitry~Ayzenberg}
\affiliation{Center for Field Theory and Particle Physics and Department of Physics, Fudan University, 200438 Shanghai, China}

\author{Cosimo~Bambi}
\email[Corresponding author: ]{bambi@fudan.edu.cn}
\affiliation{Center for Field Theory and Particle Physics and Department of Physics, Fudan University, 200438 Shanghai, China}
\affiliation{Theoretical Astrophysics, Eberhard-Karls Universit\"at T\"ubingen, 72076 T\"ubingen, Germany}

\author{Leonardo~Modesto}
\affiliation{Department of Physics, Southern University of Science and Technology (SUSTech), Shenzhen 518055, China}

\author{Sourabh~Nampalliwar}
\affiliation{Theoretical Astrophysics, Eberhard-Karls Universit\"at T\"ubingen, 72076 T\"ubingen, Germany}

\author{Yerong~Xu}
\affiliation{Center for Field Theory and Particle Physics and Department of Physics, Fudan University, 200438 Shanghai, China}

\begin{abstract}
We consider the family of singularity-free rotating black hole solutions in Einstein's conformal gravity found in Ref.~\cite{p1} and we constrain the value of the conformal parameter $L$ from the analysis of a 30~ks \textsl{NuSTAR} observation of the stellar-mass black hole in GS~1354--645 during its outburst in 2015. Our new constraint is much stronger than that found in previous work. Here we obtain $L/M < 0.12$ (99\% confidence level, statistical uncertainty only).
\end{abstract}

\maketitle


\section{Introduction}

Einstein's gravity was proposed at the end of 1915 and is still the standard framework for the description of gravitational fields and the chrono-geometrical structure of spacetime. Despite its undoubted successes to explain a large number of observational data~\cite{will}, the theory is plagued by a few but important problems that clearly point out the existence of new physics. One of these problems is the presence of spacetime singularities in physically relevant solutions of the Einstein equations. At a singularity, predictability is lost and standard physics breaks down. It is often advocated that the problem of spacetime singularities in Einstein's gravity can be fixed by the yet unknown theory of quantum gravity. However, current approaches to quantize the gravitational field do not easily solve the spacetime singularities of the classical theory.

Conformal symmetry is an appealing proposal to solve the singularity issue in Einstein's gravity~\cite{cg1,cg2,cg3,cg4,cg5,cg6,cg7}. We require that the theory is invariant under a conformal transformation of the metric tensor $g_{\mu\nu}$
\be\label{eq-tran}
g_{\mu\nu} \rar g_{\mu\nu}^* = \Omega^2 g_{\mu\nu} \, ,
\ee  
where $\Omega = \Omega (x)$ is a function of the spacetime point. Einstein's gravity is not invariant under conformal transformations, but it can be made conformally invariant (Einstein's conformal gravity) by introducing a conformal compensator field $\phi$ (dilaton). For example, a possible action is~\cite{dirac}
\be\label{eq-action}
S = - 2 \int d^4x \sqrt{-g} \left[ \phi^2 R + 6 g^{\mu\nu} 
\left(\partial_\mu \phi\right) \left(\partial_\nu \phi\right) \right] \, .
\ee
Imposing that the dilaton field transform as
\be
\phi \rar \phi^* = \Omega^{-1} \phi \, ,
\ee
the action in Eq.~(\ref{eq-action}) is invariant under the conformal transformation of the metric tensor in~(\ref{eq-tran}). Note that for $\phi = 1/\sqrt{32 \pi} = {\rm constant}$ we recover Einstein's gravity with the correct normalization.

The analogy between general covariance and conformal symmetry can illustrate how the latter can solve the problem of spacetime singularities~\cite{inter1,inter2,inter3,inter4,inter5}. In Einstein's gravity, the theory is invariant under general coordinate transformations (general covariance). If a certain quantity is singular in a coordinate system but not in another one, the singularity is not physical but just an artifact of the reference frame. We have a {\it coordinate singularity}, namely a singularity related to the coordinate system, which is not an intrinsic singularity of the spacetime. The choice of the coordinate system is arbitrary in Einstein's gravity, and therefore physical quantities cannot depend on it. As an example, we can consider the Schwarzschild spacetime in Schwarzschild coordinates. The metric is singular at the event horizon located at $r = 2M$, but the spacetime is regular there, and the singularity can be removed with a proper choice of the coordinate system. The point $r = 0$ is instead a true singularity of the spacetime. The Kretschmann scalar is invariant under general coordinate transformations and is singular at $r=0$ in every reference frame. Geodesics reaching $r = 0$ stop there in any coordinate system and we can thus say that the spacetime is geodetically incomplete at $r = 0$.

In a conformally invariant theory of gravity, the action is invariant under both general coordinate transformations and conformal transformations. If a certain quantity is singular in a reference frame but not in another one after a conformal transformation, the singularity is not physical but just an artifact of the reference frame. Now we have a {\it conformal singularity}, namely a singularity related to the choice of the conformal factor, and it is not an intrinsic singularity of the spacetime. Note that we cannot use the same mathematical tools for studying spacetime singularities in Einstein's gravity and in conformal gravity. For example, the scalar curvature and the Kretschmann scalar are not invariant under conformal transformations, and therefore they are not associated with any intrinsic property of the spacetime in conformal gravity. The study of geodesics is also different in conformal gravity because standard massive particles are not allowed as they are not compatible with conformal symmetry.

The world around us is clearly not conformally invariant. For example, in a conformally invariant theory we cannot measure lengths and time intervals, which is definitively not the case. If we want to explore the possibility that conformal invariance is a fundamental symmetry in Nature, we must admit that around us such a symmetry is broken. If conformal invariance is spontaneously broken, Nature has selected one of the possible vacua. The problem of spacetime singularities can be solved postulating that Nature can only select a physical vacuum in the class of singularity-free metrics~\cite{inter1,inter2,inter3,inter4,inter5}.

In Ref.~\cite{p1}, we found a singularity-free, exact, rotating black hole solution of a large family of conformally invariant theories of gravity. In Boyer-Lindquist coordinates, the line element reads\footnote{We employ units in which $G_{\rm N} = c = 1$ and a metric with signature $(-+++)$.}
\be\label{eq-ds}
ds^2 = \left( 1 + \frac{L^2}{\Sigma} \right)^2 ds^2_{\rm Kerr} 
\ee
where $ds^2_{\rm Kerr}$ is the line element of the Kerr metric
\be
ds^2_{\rm Kerr} &=&  - \left( 1 - \frac{2 M r}{\Sigma} \right) \, dt^2
- \frac{4 M a r \sin^2\theta}{\Sigma} \, dt \, d\phi \nonumber\\
&& + \frac{\Sigma}{\Delta} \, dr^2 + \Sigma \, d\theta^2
\nonumber\\ &&
+ \left(r^2 + a^2 
+ \frac{2 M a^2 r \sin^2\theta}{\Sigma}\right) \, \sin^2\theta \, d\phi^2  \, ,
\ee
$\Sigma = r^2 + a^2 \cos^2\theta$, $M$ is the black hole mass, $a = J/M$ is the specific spin, $J$ is the black hole spin angular momentum, and $L$ is the conformal parameter. The theory does not give any indication about the value of $L$, but it is natural to expect it is either of the order of the Planck length, $L \sim L_{\rm Pl} \sim 10^{-33}$~cm, or of the order of the black hole mass, $L \sim M$, as these are the only two scales already present in the system. The scenario with $L \sim L_{\rm Pl}$ likely cannot be tested with astrophysical observations. In what follows, we assume $L \sim M$.

Astrophysical observations of black holes can constrain the conformal parameter $L$ because now we are in a broken phase and therefore reference frames that differ by a conformal transformation are not equivalent as they are during the symmetric phase. In Ref.~\cite{p2}, we studied the iron K$\alpha$ line expected in the reflection spectrum of accretion disks around singularity-free black holes and how the line shape is affected by the conformal parameter $L$. For fast-rotating black holes, as $L$ increases the radius of the innermost stable circular orbit (ISCO) increases as well, and this has clear observational implications in the iron line shape. In particular, for sufficiently large values of $L/M$, we cannot have a very broad iron line. Since we have observations of broad iron lines in the X-ray spectrum of black holes, we can obtain the constraint $L/M < 1.2$. In Ref.~\cite{p2b}, we employed a modified version of the reflection model {\sc relxill\_nk}~\cite{apj,0707,564,339,shenzhen}, and we analyzed the 2014 observations of \textsl{NuSTAR} and \textsl{Swift} of the supermassive black holes in 1H0707--495, obtaining the constraint $L/M < 0.45$ (90\% confidence level). However, 1H0707--495 is quite a controversial source, not fully understood at the moment, and therefore the constraint is not robust.

In the present paper, we analyze one of the three \textsl{NuSTAR} observations of the 2015 outburst of the X-ray binary GS~1354--645. The spectrum of this source is clearly reflection dominated and our analysis, in agreement with previous studies, suggests that the inner edge of the disk is extremely close to the compact object. We are thus able to find very strong constraints on the spin parameter $a_* = a/M$ and the conformal parameter $L$. Our result is $a_* > 0.985$ and $L/M < 0.12$ (99\% confidence level).


\section{X-ray reflection spectroscopy \label{s-x}}

The standard framework for the description of accreting black holes is the disk-corona model (see, for instance, Ref.~\cite{rev}). The central black hole is surrounded by a geometrically thin and optically thick accretion disk. At every point of the disk, the emission is like that of a blackbody, and the accretion disk has a multi-temperature blackbody spectrum. The temperature of the disk scales as $M^{-1/4}$ and the emission is generally peaked in the soft X-ray band (0.1-1~keV) for stellar-mass black holes and in the optical/UV band (1-10~eV) for the supermassive ones. The term ``corona'' is used to indicate a hotter ($\sim 100$~keV), usually compact and optically thin, medium close to the black hole. Its exact nature and morphology is currently not well understood. In the so-called lamppost geometry, the corona is a point-like source along the spin axis of the black hole, and it may be interpreted as the base of a jet. In the so-called sandwich geometry, the corona is the atmosphere covering the accretion disk.

Thermal photons from the accretion disk can have inverse Compton scattering off free electrons in the corona. Such a process produces a spectrum that can be approximated with a power-law component ($\sim E^{-\Gamma}$, where $\Gamma \approx 1$-3 is called the photon index) with an energy cut-off $E_{\rm cut} \sim 100$~keV. These photons can illuminate the accretion disk, producing a reflection component with some emission lines. X-ray reflection spectroscopy refers to the study of this reflection component.

The reflection spectrum is usually characterized by the iron K$\alpha$ line complex around 6~keV and by the Compton hump at 10-30~keV. In the rest-frame of the emitting medium, the iron K$\alpha$ line is a very narrow feature at 6.4~keV in the case of neutral or weakly ionized iron, and it shifts up to 6.97~keV in the case of H-like iron ions. On the contrary, the iron K$\alpha$ line detected in the reflection spectrum of accreting black holes is very broad and skewed, as a result of relativistic effects occurring in the strong gravity region around the compact object (for a review, see, for instance, Ref.~\cite{rmp}). While the iron K$\alpha$ line is usually the most informative feature about the spacetime geometry around the black hole, any accurate measurement of the black hole metric necessarily requires fitting the whole reflection spectrum, not only the iron line.

Assuming that the spacetime metric around astrophysical black holes is described by the Kerr solution of Einstein's gravity, X-ray reflection spectroscopy can measure black hole spins~\cite{k1,k2,k3}. There are currently about 10~stellar-mass black holes and about 30~supermassive black holes with a spin measurement obtained through analyzing their reflection spectrum. More recently, there has been an increasing interest in the possibility of using X-ray reflection spectroscopy to test the predictions of Einstein's gravity in the strong field regime~\cite{nk1,nk2,nk3,nk4,nk5,nk6,nk7,nk8,nk9}.

As of now, the {\sc relxill} package is the most advanced model for the description of the reflection spectrum of accretion disks in the Kerr metric~\cite{re1,re2,re3,re4}. In Ref.~\cite{apj}, we presented {\sc relxill\_nk}, which is the extension of {\sc relxill} to non-Kerr spacetimes. The model employs the Johannsen metric~\cite{j} and has been used to test the Kerr metric with the supermassive black holes in 1H0707--495~\cite{0707} and Ark~564~\cite{564}, and the stellar mass black hole in GX~339-4~\cite{339} (for a summary of current constraints, see Ref.~\cite{shenzhen}). In the present paper, we will use a modified version of {\sc relxill\_nk} employing the black hole metric in Eq.~(\ref{eq-ds}) in order to constrain the conformal parameter $L$. We will analyze a 30~ks \textsl{NuSTAR} observation of the stellar-mass black hole in GS~1354--645. More details on {\sc relxill\_nk}, its parameters, and the tests performed to validate this model, see Ref.~\cite{apj}.

\begin{figure}[t]
\begin{center}
\includegraphics[type=pdf,ext=.pdf,read=.pdf,width=8.5cm]{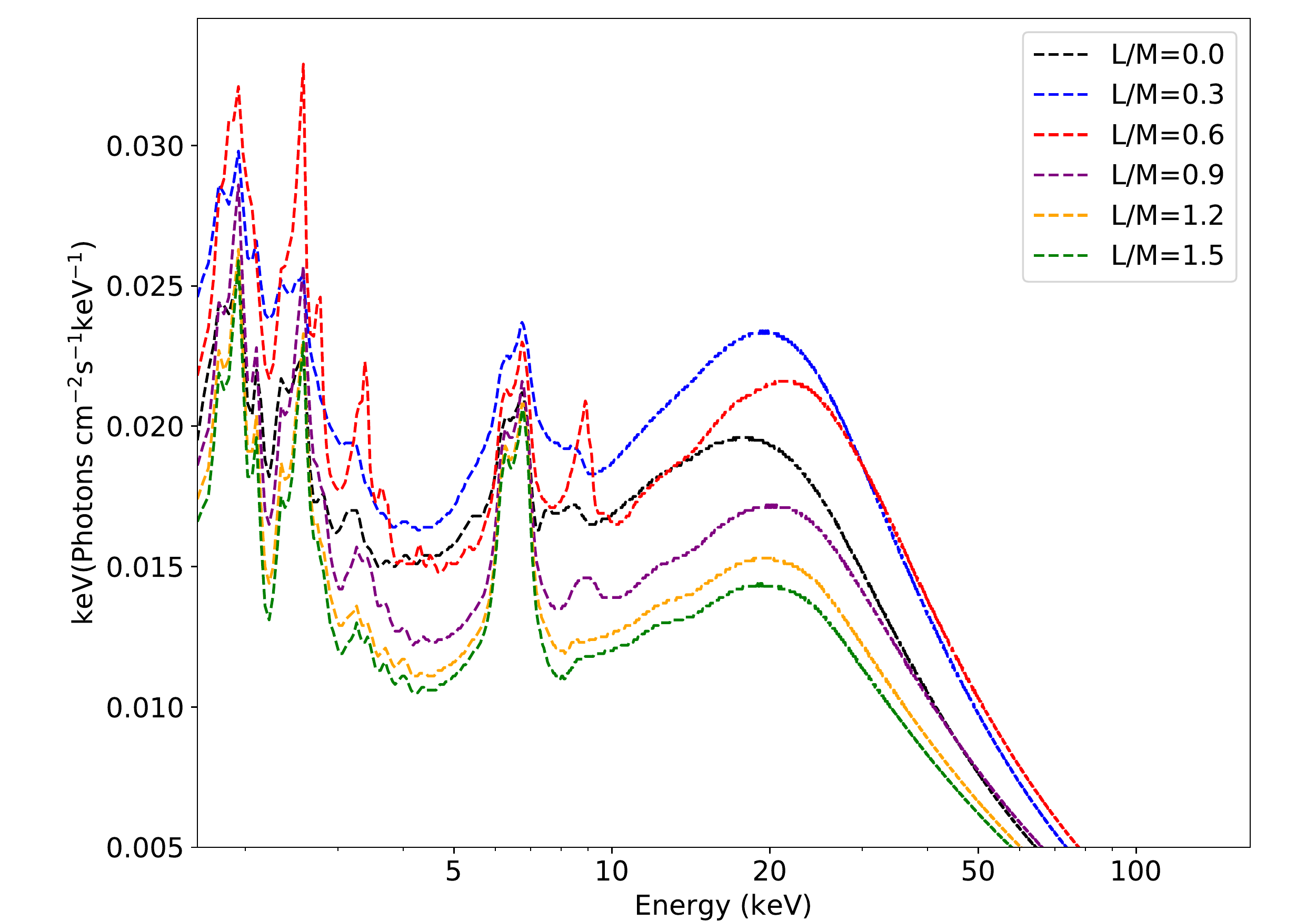}
\end{center}
\vspace{-0.4cm}
\caption{Impact of the conformal parameter $L$ on the reflection spectrum of a thin accretion disk around a black holes in conformal gravity. The other model parameters are: spin $a_* = 0.998$, inclination angle $i = 75$~deg, emissivity indices $q_{\rm in} = q_{\rm out} = 3$, iron abundance $A_{\rm Fe} = 1$ (i.e. Solar iron abundance), ionization parameter $\log\xi = 3.1$, photon index of the radiation illuminating the disk $\Gamma = 2$. \label{f-sp}}
\end{figure}

As shown in Fig.~\ref{f-sp}, the conformal parameter $L$ has a significant impact on the reflection spectrum and it is thus clear that by fitting the X-ray spectrum of a black hole we can constrain $L$. Here, from the analysis of a 30~ks \textsl{NuSTAR} observation of the stellar-mass black hole in GS~1354--645, we will obtain quite a stringent constraint on $L/M$.

\section{Observations and data reduction \label{s-data}}

GS~1354--645 is a low-mass X-ray binary. The source was discovered by the Japanese X-ray mission \textsl{Ginga} in 1987~\cite{makino}. The measurement of the mass of the black hole is $M_{\rm BH} \ge 7.6 \pm 0.7$~$M_\odot$~\cite{casares}. The distance to the source is poorly constrained and ranges between 25 and 61~kpc~\cite{casares}. The last outburst of GS~1354--645 was in 2015~\cite{miller}. There are three \textsl{NuSTAR} observations on archive of the 2015 outburst. The first observation was on June~13 for about 24~ks (Obs. ID~90101006002). The second observation was on July~11 for about 30~ks (Obs. ID~90101006004). Lastly, \textsl{NuSTAR} observed GS~1354--645 on August~6 for about 35~ks (Obs. ID~90101006006). The first two observations were first studied in Ref.~\cite{gs}, while there is currently no publication reporting the analysis of the third one.

In the present paper, we only consider the second observation of July~11. The first observation is indeed unsuitable for our test, because the inner edge of the accretion disk is truncated at a radius much larger than the ISCO~\cite{gs}, and therefore it does not permit us to probe the strong gravity region around the black hole. On the contrary, previous studies of the second observation suggest that the inner edge of the accretion disk was extremely close to the black hole on July~11~\cite{gs,yerong}. This helps to constrain the conformal parameter $L$ because as $L$ increases it is not possible to have very small ISCO radii~\cite{p2}.

We processed the data from both the FPMA and FPMB instruments using {\it nupipeline} v0.4.5 with the standard filtering criteria and the \textsl{NuSTAR} CALDB version 20171002. We used the {\it nuproducts} routine to extract source spectra, responses, and background spectra. For the source, we chose a circular region of radius 148~arc seconds. For the background, we chose a circular region of radius 148~arc seconds on the same chip. All spectra were binned to a minimum of 30~counts before analysis to ensure the validity of the $\chi^2$ fit statistics.


\section{Spectral analysis \label{s-ana}}

We analyze the \textsl{NuSTAR} observation of July~11 using Xspec v12.9.1~\cite{arnaud}.

\begin{figure*}[t]
\begin{center}
\includegraphics[type=pdf,ext=.pdf,read=.pdf,width=7.5cm]{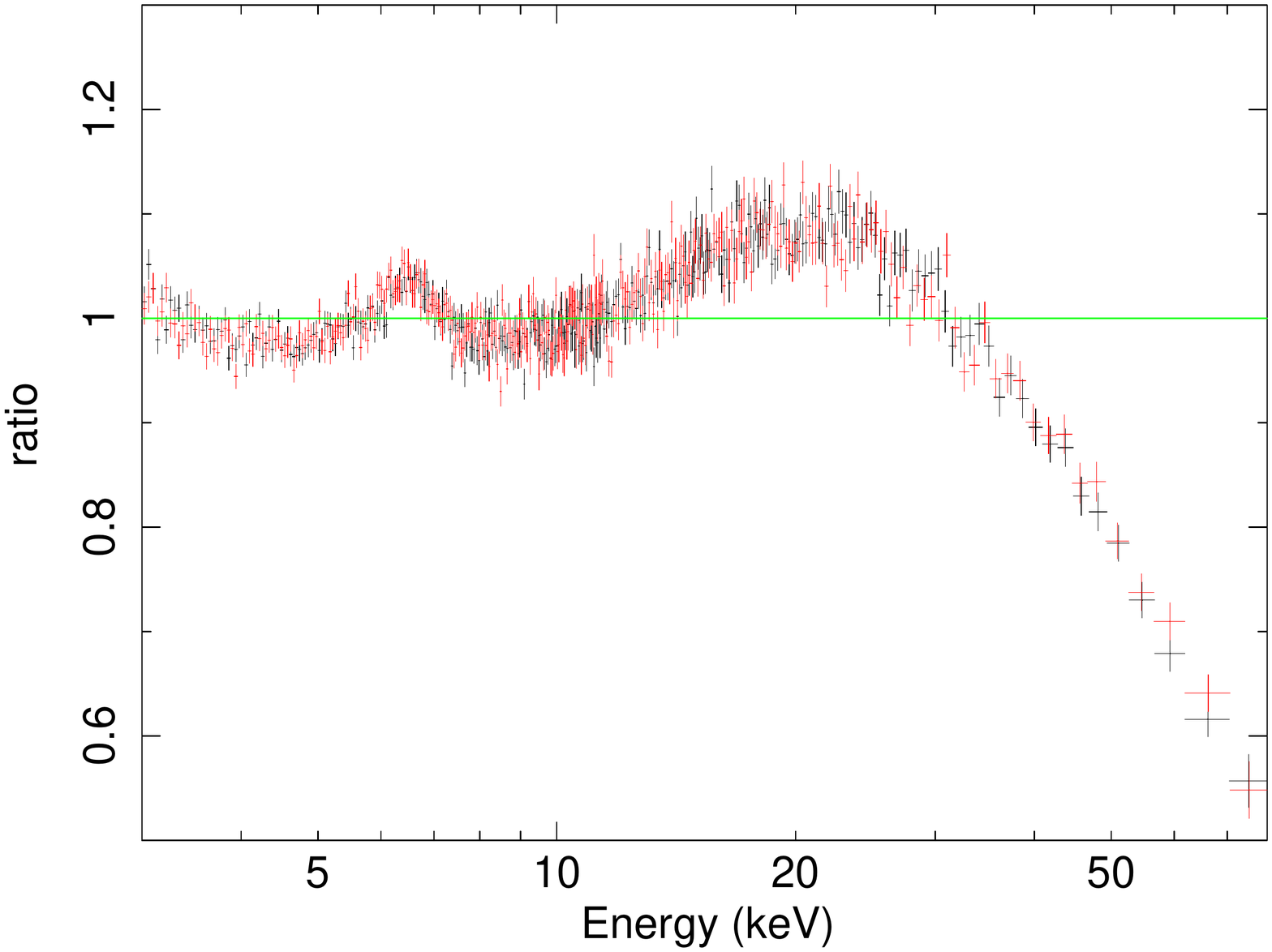}
\includegraphics[type=pdf,ext=.pdf,read=.pdf,width=7.5cm]{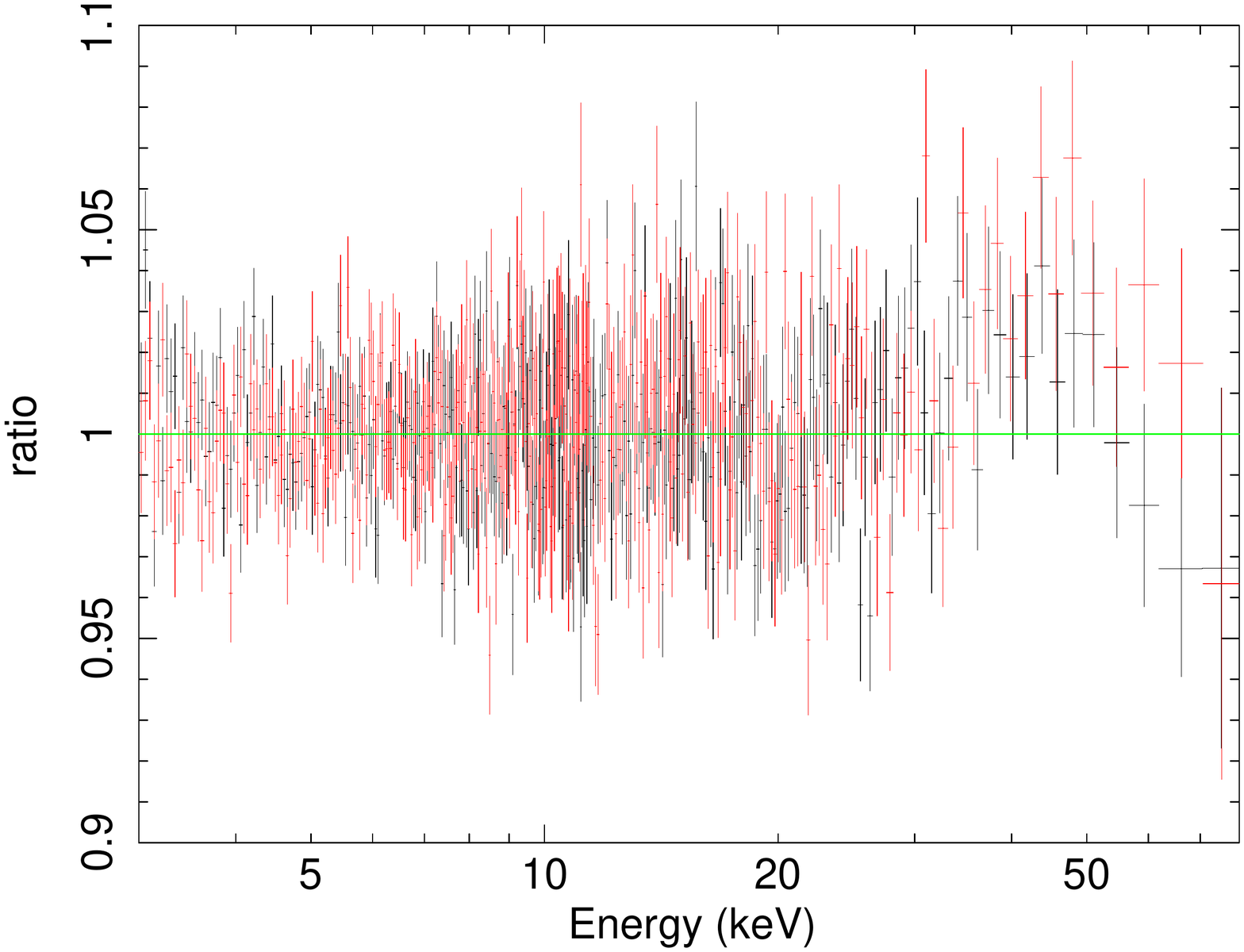}
\end{center}
\vspace{-0.5cm}
\caption{Data-to-model ratios of the 30~ks \textsl{NuSTAR} observation of July~11 of the stellar-mass black hole in GS~1354--645 employing a simple power-law model (left panel) and for a power-law and reflection model (right panel). Black crosses are used for FPMA and red crosses are used for FPMB. The data have been rebinned for plotting purposes only. \label{f-ratio}}
\end{figure*}

For the first fit, we consider an absorbed power-law model: {\sc tbabs*powerlaw}. {\sc tbabs} describes the Galactic absorption~\citep{wilms} and we fix the galactic column density to $N_{\rm H} = 0.7 \cdot 10^{22}$~cm$^{-2}$, which we obtain from the HEASARC column density tool, based on~\cite{dickey90}. The left panel in Fig.~\ref{f-ratio} shows the data to the best-fit model ratio. The spectrum of the source is clearly reflection dominated, and we see a broad iron line around 6~keV and a Compton hump at 10-30~keV.

For the second fit, we add a reflection component. Our Xspec model is {\sc tbabs*relxill\_nk}. Note that {\sc relxill\_nk} includes both the power-law component from the corona and the reflection component from the disk. The best-fit values of the model parameters are reported in Tab.~\ref{t-fit}. The right panel in Fig.~\ref{f-ratio} shows the data to the best-fit model ratio. In agreement with previous studies~\cite{gs}, we find a good fit when assuming a broken power-law for the emissivity profile of the disk, and that the inner emissivity index $q_{\rm in}$ is very high and the outer emissivity index $q_{\rm out}$ is low. Such a high value for $q_{\rm in}$ can be interpreted with a lamppost corona very close to the black hole~\cite{re1}, implying that most of the radiation is emitted from the region near the inner edge of the accretion disk. The very low value of $q_{\rm out}$ may instead be interpreted with the fact that the emission at larger radii is so low that it cannot be easily constrained.

The inner edge of the accretion disk seems to be very close to the black hole, implying both a very high black hole spin and a very low value of the conformal parameter $L$. We have repeated our analysis when relaxing the common assumption that the inner edge of the accretion disk is at the ISCO radius, without finding any difference. This is to be expected since the data require an inner edge so close to the black hole that values larger than the ISCO are disfavored. We note that the best-fit values of our model parameters is consistent with the study of Ref.~\cite{gs}.

\begin{figure}[t]
\begin{center}
\includegraphics[type=pdf,ext=.pdf,read=.pdf,width=8.5cm]{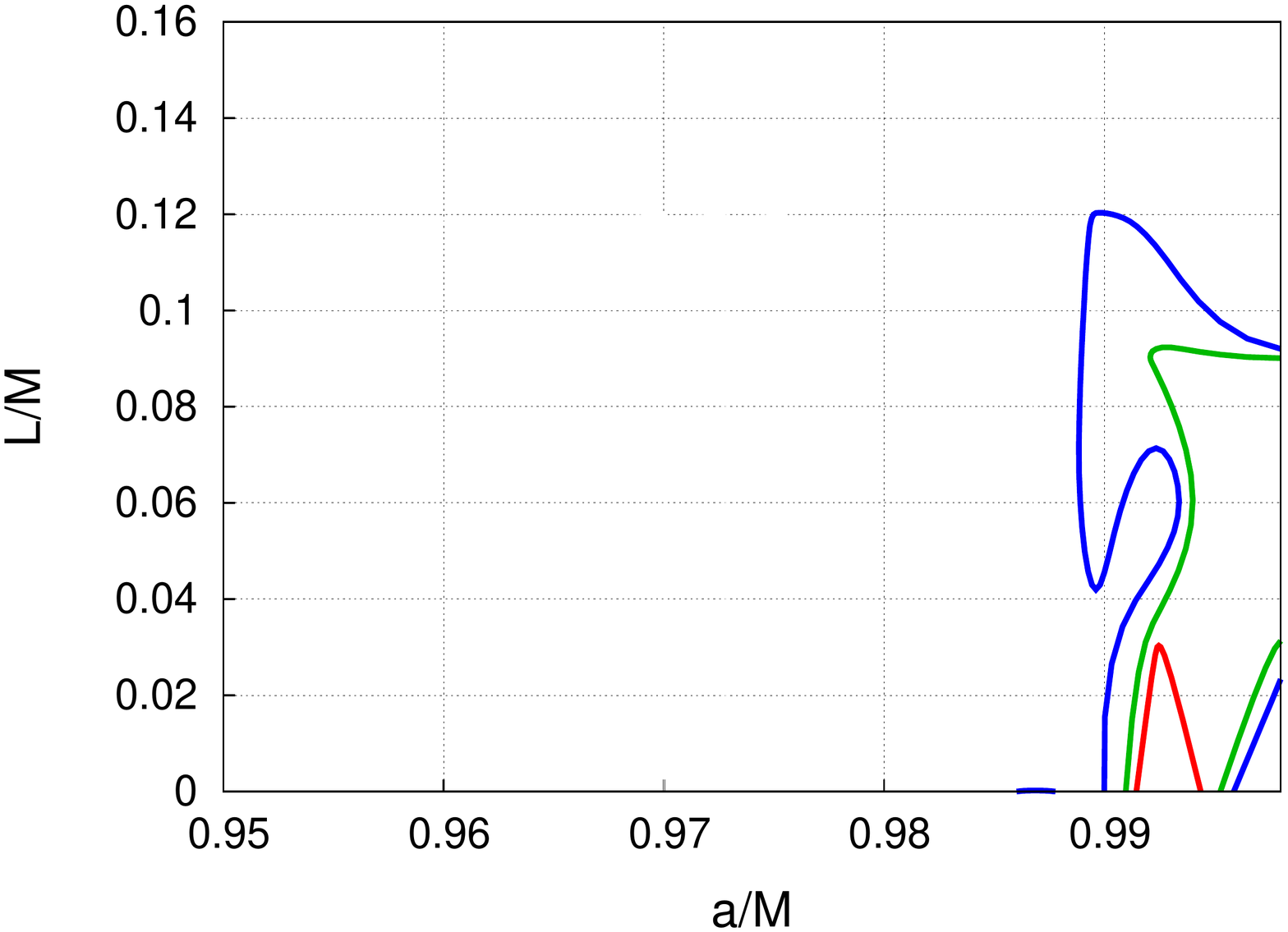}
\end{center}
\vspace{-0.9cm}
\caption{Constraints on the specific spin $a$ and the conformal parameter $L$ from the 30~ks \textsl{NuSTAR} observation of July~11 of the stellar-mass black hole in GS~1354--645. The red, green, and blue lines indicate, respectively, the 68\%, 90\%, and 99\% confidence level contours for two relevant parameters. \label{f-plot}}
\end{figure}

Our constraints on the black hole spin and the conformal factor $L$ are shown in Fig.~\ref{f-plot}. The red, green, and blue lines indicate, respectively, the 68\%, 90\%, and 99\% confidence level contours for two relevant parameters. Within 99\% confidence level, our measurements are
\be
a/M > 0.985 \, \qquad
L/M < 0.12 \, .
\ee
Note that our constraints only include the statistical uncertainty. Systematic uncertainties are more difficult to estimate. An important assumption in our model is that the accretion disk is geometrically thin. This is usually thought to be a good approximation for sources accreting between 5\% and 20\% of their Eddington limit, see Refs.~\cite{penna,ssp}. Unfortunately, the distance and the mass of the black hole in GS~1354--645 are currently poorly constrained~\cite{casares}. On July~11, the luminosity of the source was $L/L_{\rm Edd} \le 0.53$, which is consistent with the 5-20\% range but it also allows higher and lower luminosities. There are several other approximations in the model, including the calculations of the reflection spectrum at the emission point (see, for instance, \cite{339} and references therein). Despite that, it is clear that the July~11 observation of \textsl{NuSTAR} is reflection dominated and requires an inner edge of the accretion disk very close to the black hole, and this permits us to constrain the conformal factor $L$ to a value smaller than the gravitational radius of the system $M$.

\begin{table*}[t]
\vspace{0.5cm}
\centering
\begin{tabular}{lcccc}
\hline\hline
Model parameter & \hspace{0.5cm} & Value & \hspace{0.5cm} & Description and units \\
\hline
{\sc tbabs} &&&& \\
$N_{\rm H}$ && $0.7 \cdot 10^{22 \, \star}$ && Column density (cm$^{-2}$) \\
\hline
{\sc relxill\_nk} &&&& \\
$a$ && $0.9924^{+0.014}_{-0.015}$ && Specific spin ($M$) \\
$L$ && $< 0.024$ && Conformal parameter ($M$) \\
$R_{\rm in}$ && $1^\star$ && Inner radius ($R_{\rm ISCO}$)\\
$i$ && $75.7^{+0.8}_{-1.2}$ && Disk inclination angle (deg) \\
$q_{\rm in}$ && $8.79^{+0.49}_{-0.24}$ && Inner emissivity index \\
$q_{\rm out}$ && $0.48^{+0.05}_{-0.05}$ && Outer emissivity index \\
$R_{\rm br}$ && $5.2^{+0.7}_{-0.4}$ && Breaking radius ($M$) \\
$\Gamma$ && $1.658^{+0.010}_{-0.008}$ && Photon index of the power-law component \\
$\log\xi$ && $2.38^{+0.05}_{-0.03}$ && Ionization parameter (erg~cm~s$^{-1}$)\\
$A_{\rm Fe}$ && $0.51^{+0.04}$ && Iron abundance (Solar) \\
$E_{\rm cut}$ && $161^{+10}_{-10}$ && High energy cut-off (keV)\\
$R$ && $1.30^{+0.03}_{-0.07}$ && Reflection fraction \\
\hline
$\chi^2$/dof && 2886/2733 && \\
 && = 1.061 &&\\
\hline\hline
\end{tabular}
\vspace{0.3cm}
\caption{Summary of the best-fit values for the model {\sc tbabs*relxill\_nk}. The reported uncertainties correspond to the 90\% confidence level for one relevant parameter. $^\star$ indicates that the parameter is frozen in the fit. \label{t-fit}}
\end{table*}

\section{Concluding remarks \label{s-con}}

In the present paper, we have tested the singularity-free black hole metrics found in Ref.~\cite{p1}. These spacetimes are exact solutions in a large family of conformal theories of gravity and are characterized by the conformal factor $L$. For $L=0$, we recover the singular Kerr metric of Einstein's gravity. While there are no indications from the theory on the value of $L$, it is natural to expect that $L$ is either of the order of the Planck length or of the order of the mass of the black hole $M$, as these are the only two scales already present in the system. Here we have considered the second scenario, which is likely the only one with astrophysical implications.

We have applied a modified version of our recent X-ray reflection model {\sc relxill\_nk}~\cite{apj} to a 30~ks \textsl{NuSTAR} observation of the stellar-mass black hole in GS~1354--645 during its 2015 outburst. The spectrum of the source is clearly reflection dominated and our analysis, in agreement with previous studies, finds that the inner edge of the accretion disk is extremely close to the compact object, which is particularly useful for placing a strong constraint on the conformal parameter $L$. Our measurements of the black hole spin and of the conformal parameter are, respectively, $a_* > 0.985$ and $L/M < 0.12$ within 99\% confidence level and only including the statistical uncertainty.


\begin{acknowledgments}
This work was supported by the National Natural Science Foundation of China (NSFC), Grant No.~U1531117, and Fudan University, Grant No.~IDH1512060. A.B.A. also acknowledges the support from the Shanghai Government Scholarship (SGS). C.B. also acknowledges support from the Alexander von Humboldt Foundation. S.N. also acknowledges support from the Excellence Initiative at Eberhard-Karls Universit\"at T\"ubingen.
\end{acknowledgments}


\end{document}